# Machine learning-based decentralized TDMA for VLC IoT networks

Armin Makvandi [1] and Yousef Seifi Kavian [1,*]

[1] Department of Electrical Engineering, Faculty of Engineering, Shahid Chamran University of Ahvaz, Iran

[*] Corresponding author: y.s.kavian@scu.ac.ir

*Abstract*—In this paper, a machine learning-based decentralized time division multiple access (TDMA) algorithm for visible light communication (VLC) Internet of Things (IoT) networks is proposed. The proposed algorithm is based on Q-learning, a reinforcement learning algorithm. This paper considers a decentralized condition in which there is no coordinator node for sending synchronization frames and assigning transmission time slots to other nodes. The proposed algorithm uses a decentralized manner for synchronization, and each node uses the Q-learning algorithm to find the optimal transmission time slot for sending data without collisions. The proposed algorithm is implemented on a VLC hardware system, which had been designed and implemented in our laboratory. Average reward, convergence time, goodput, average delay, and data packet size are evaluated parameters. The results show that the proposed algorithm converges quickly and provides collision-free decentralized TDMA for the network. The proposed algorithm is compared with carrier-sense multiple access with collision avoidance (CSMA/CA) algorithm as a potential selection for decentralized VLC IoT networks. The results show that the proposed algorithm provides up to 61% more goodput and up to 49% less average delay than CSMA/CA.

## 1 INTRODUCTION

In recent years, the use of wireless communication devices has increased significantly, causing the radio frequency (RF) spectrum to be saturated in the future. As a result, complementary technologies for this purpose are required. Visible light communication (VLC) has unique characteristics, which make it a good choice for wireless networks. This technology uses the light-emitting diode (LED), providing illumination and communication simultaneously. Free-license wide bandwidth alongside a high data rate makes it a key technology for sixth-generation (6G) wireless communication [1]. Low power consumption and high security make it a good choice for Internet of Things (IoT) applications [2]. Some other applications of this technology include smart lighting [3], underwater communications [4], location-based services [5], and vehicular communications [6]. VLC is safe for human health and does not interfere with RF devices, making it a good choice for use in sensitive environments such as hospitals [7]. However, weak non-line-of-sight (NLOS) communication and the short-distance range of VLC compared to RF limit the use of this technology for specific applications. Therefore, VLC is an excellent complement to RF for 6G wireless communication and also can be used in hybrid RF-VLC networks [8].

Recently, several VLC hardware systems for IoT applications have been designed [9, 10]. One of the most critical aspects of an IoT system is the network performance, in which several IoT devices should share a channel and send their data to each other or gateway. If multiple devices send data to one device at the same time, the data collision will occur, degrading the performance of the network. Multiple access protocols are responsible for controlling the access of multiple users to the channel. Depending on the wireless network characteristics and desired scenarios, different types of multiple access algorithms can be used. Different multiple access algorithms have been proposed for VLC networks, but most have limitations in IoT applications. In [11], an algorithm based on ALOHA was proposed for VLC networks. In this protocol, each device that has data for transmission immediately sends it. This algorithm is simple and decentralized, making it suitable for IoT applications. However, due to the lack of a strategy to avoid collisions, increasing the number of nodes and network load degrades the network performance significantly. The IEEE 802.15.7 standard [12] proposed an algorithm based on carrier-sense multiple access with collision avoidance (CSMA/CA) for VLC networks. In this algorithm, each device that has data for transmission waits for a random time and optionally checks the channel status. If the channel is clear, the device sends its data. If the channel is busy, the node postpones its transmission and does a retransmission procedure. Recently, the performance of this algorithm has been investigated in VLC networks [13, 14], and optimization solutions have been proposed [15, 16]. The simplicity and decentralized nature of this algorithm make it suitable for IoT applications. Furthermore, its collision avoidance strategy improves the network performance compared to ALOHA in VLC IoT networks [17]. However, its contention-based nature degrades its performance in congested networks. Another type of multiple access algorithms are collision-free protocols, providing a higher quality of service (QoS) than contention-based algorithms, but have other challenges in VLC IoT networks. Several types of these algorithms have been proposed for VLC networks. Li et al. [18] proposed a code-division multiple access (CDMA)-based algorithm for VLC networks. In this algorithm, each node is assigned a unique code that modulates the transmitted signal. These codes are orthogonal and do not interfere with each other. The receiver node decodes the signals by matching them to the corresponding codes. Therefore, multiple nodes can transmit and receive data over the same VLC channel simultaneously. This algorithm is spectrally efficient, decentralized, and provides parallel Communication. However, the high level of complexity of this algorithm requires advanced signal processing, increasing the complexity and cost of devices. As a result, this algorithm is not suitable for VLC IoT applications, in which simplicity and low-cost implementation are required. Orthogonal frequency-division multiple access (OFDMA) [19] is another type of collision-free algorithm for VLC networks. This method is a modulation and access technique used in VLC networks, dividing the available optical spectrum into multiple orthogonal subcarriers, each carrying a different part of the data. This technique is decentralized but is not spectral efficient enough. Furthermore, a high level of signal processing is required, making it inappropriate for VLC IoT applications, in which the system should be simple and low-cost. Marshoud et al. [20] proposed optical non-orthogonal multiple access (O-NOMA) for VLC networks. In this algorithm, data is encoded using different power levels, and the receiver decodes the data based on the received power levels. The non-orthogonal nature of this algorithm requires advanced signal processing techniques at the receiver to manage interference. As a result, this complex algorithm is not suitable for simple and low-cost VLC IoT applications.

One of the most popular contention-free multiple access algorithms, which has been used in VLC networks [21, 22], is time division multiple access (TDMA). In this algorithm, each user should transmit its data in a predefined time slot, ensuring a collision-free network. Assigning time slots to different users in the network and precise synchronization between them is challenging. In the traditional type of TDMA, a coordinator communicates with all nodes to provide synchronization and assign different transmission time slots to all nodes. In some IoT applications, there is no coordinator in the network. Therefore, the traditional TDMA cannot be used, and a decentralized TDMA is required. Mao et al. [23] proposed a decentralized algorithm for VLC networks combining TDMA and CDMA. This algorithm operates in two phases: neighbor discovery and data transmission. During the neighbor discovery phase, a node that wants to join the network listens to the channel to receive information from its neighbors and obtain the time slots and optical orthogonal codes of its neighbors. The node computes the signal-to-noise ratio (SNR) to estimate the channel quality and determine whether a node is within the communication range. Based on the estimated SNR, the node can update its neighbor set and choose an unused time slot to transmit data. This algorithm is decentralized but suffers from a long delay time during the recovery phase. Moreover, the use of CDMA increases the required signal processing, complexity, and hardware cost, which is not suitable for practical VLC IoT applications. Boukhalfa et al. [24] proposed a distributed TDMA algorithm for VLC networks. In this algorithm, when nodes compete for a time slot, they generate a burst signal by drawing a random binary transmission key. If a node senses a competing transmission during its listening period, it withdraws from the competition. If no competing transmission is sensed, the node continues to the next digit of its transmission key. As the competition progresses, there are fewer and fewer nodes competing for a time slot from one mini-slot to the next, reducing the chances of access collisions. The slot scheduling table is updated each time a node sends a packet, and the frame information field in the packet specifies the status of slots. Nodes use this information to select available slots in the next frame. This dynamic distributed algorithm is beneficial in some VLC networks, but its high overhead reduces the network performance. Furthermore, this algorithm is not collision-free because nodes compete for time slots continuously, decreasing the algorithm's reliability in critical VLC IoT applications. Huang et al. [25] proposed a distributed multi-slot TDMA algorithm for VLC networks. In this algorithm, each node that needs a time slot to send its data generates a request packet with information about out-neighbors, selects a time slot randomly from the reserve slots, and enters the waiting phase. Responding nodes assign available time slots to the requesting node using a slot-occupation query table based on in-neighbor slot-occupation data. If available, a slot is randomly chosen and sent; otherwise, a specific assignment scheme is executed. During this time, the requesting node waits for slot assignment packets from out-neighbors. Upon receiving one slot assignment packet, it enters a stable state, indicating a successful request. This decentralized algorithm has low overhead and high throughput and provides a collision-free network using a dynamical collision detection and avoidance method. However, this dynamic approach increases delay and energy consumption in the network, weakening this algorithm for VLC IoT applications.

To the best of our knowledge, previous decentralized TDMA algorithms in VLC networks have limitations for IoT applications. In this paper, a machine learning-based decentralized TDMA algorithm for VLC IoT networks is proposed. The proposed algorithm is based on Q-learning, a reinforcement learning algorithm. This paper considers a decentralized condition in which there is no coordinator node for sending synchronization frames and assigning transmission time slots to other nodes. The proposed algorithm uses a decentralized manner for synchronization, and each node uses the Q-learning algorithm to find the optimal transmission time slot for sending data without collisions. The proposed algorithm is implemented on a VLC hardware system [9], which had been designed and implemented in our laboratory. For evaluation, average reward, convergence time, goodput, average delay, and data packet size are considered. The results show that the proposed algorithm converges quickly and provides collision-free decentralized TDMA for the network. Comparison of the proposed algorithm with CSMA/CA, as a potential selection for decentralized VLC IoT networks, shows the superiority of the proposed algorithm in terms of goodput and average delay.

The main contributions of this paper are summarized as follows:
- This paper proposes a machine learning-based decentralized TDMA algorithm for VLC IoT networks to overcome the limitations of previous decentralized TDMA algorithms for this purpose.
- The proposed algorithm is implemented on a VLC hardware system, and the results show that it converges quickly and provides collision-free decentralized TDMA for VLC IoT networks.
- Comparison of the proposed algorithm with CSMA/CA, as a potential selection for decentralized VLC IoT networks, shows the superiority of the proposed algorithm in terms of goodput and average delay.

The rest of this paper is organized as follows. In Section 2, the proposed algorithm is described. In Section 3, the experimental results show the performance of the proposed algorithm and its comparison with CSMA/CA in the VLC IoT network. Finally, the conclusion and future works are given in Section 4.

## 2 PROPOSED ALGORITHM

### 2.1 Reinforcement learning

Reinforcement learning (RL) is an interactive type of machine learning [26]. In this approach, an agent uses trial and error and interacts with the environment. The agent takes actions and receives corresponding rewards from the environment. The goal of the agent is to maximize cumulative reward. In Fig. 1, the procedure of RL in a Markov decision process (MDP) is shown. The agent follows this procedure in each time slot ($t$). $R$ is the reward, $S$ is a set of possible states, and $A$ is a set of possible actions. At each time step, the agent senses the current state of the environment $S_t \in S$ and selects an action $A_t \in A(S_t)$, where $A(S_t)$ is the set of available actions in state $S_t$. Taking an action changes the state to $S_{t+1}$ and the agent receives a reward $R_{t+1} \in R$. $R_t$ and $S_t$ depend only on the previous state and action, which is the nature of an MDP. Selecting the action is done according to current policy ($\pi$), which is the probability of selecting action $A_t = a$ when the agent is in state $S_t = s$.

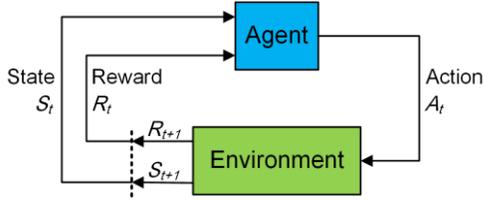

Fig. 1. Reinforcement learning process.

The agent should obtain optimal policy ($\pi^*$), requiring solving the Bellman optimality equation. This equation for state-value function is derived as

$$V_*(s) = \max_{a \in A} \sum_{s' \in S} P(s'|s, a)[R + \gamma V_*(s')] \quad (1)$$

where $V_*(s)$ represents the optimal state-value function, $P(s'|s, a)$ is the state-transition probability function, $V_*(s')$ is the state-value function for the next state $S_{t+1}$, $R$ is the reward function, and $\gamma$ is the discount factor in which $\gamma \in (0, 1)$. The Bellman optimality equation for the action-value function is derived as

$$q_*(s, a) = \sum_{s' \in S} P(s'|s, a)\left[R + \gamma \max_{a' \in A} q_*(s', a')\right] \quad (2)$$

where $q_*(s)$ represents the optimal action-value function. In practical scenarios, such as channel sharing in this paper, the dynamic of the environment $P(s'|s, a)$ is not well known by the agent. As a result, in these scenarios, for solving the equation, approximation methods are required.

### 2.2 Q-learning

One of the most powerful reinforcement learning algorithms is Q-learning [27]. This algorithm does not need the environment to be modeled and finds the approximate solution to the Bellman optimality equation. This algorithm is off-policy, which means the agent approximates the optimal action-value function ($q_*$) independent of the policy. This algorithm uses a Q-function, which indicates the quality of a specific action at a specific state. Q-values are updated as follows:

$$Q(S_t, A_t) \leftarrow Q(S_t, A_t) + \alpha\left[R_{t+1} + \gamma \max_a Q(S_{t+1}, a) - Q(S_t, A_t)\right] \quad (3)$$

where $\alpha$ is the learning rate in which $\alpha \in (0, 1]$.

### 2.3 Exploration and exploitation

In reinforcement learning, the agent should explore the environment to find the best action. On the other hand, exploitation is used to utilize the experience already available. There are several methods to provide a good balance between exploration and exploitation. One of these methods is greedy selection, in which the agent always selects the action with the highest action value. This approach does not ensure a good balance between exploration and exploitation because it does not control the exploration time. Another method is $\epsilon$-greedy selection. In this method, the action with the highest action value is selected with probability $\epsilon$, and other actions are selected with probability 1-$\epsilon$. A High value of $\epsilon$ causes the agent to explore frequently, and a lower value of $\epsilon$ causes the agent to exploit its current knowledge primarily. To balance between exploration and exploitation and ensure convergence to optimal policy, a decay rate for $\epsilon$ can be used. Initially, a high value for $\epsilon$ is considered to ensure sufficient exploration and earn the required environmental knowledge. Over time, the decay rate decreases the $\epsilon$ to ensure that the agent focuses more on exploitation as it has more knowledge about the environment.

### 2.4 Proposed algorithm

In this paper, a decentralized condition is considered, in which there is no coordinator for sending synchronization frames and assigning transmission time slots to other nodes. Therefore, the traditional TDMA cannot be used. The proposed algorithm uses a decentralized manner for synchronization. For this purpose, when a node wants to join the network, it broadcasts a predefined synchronization frame to other nodes. All nodes in the network that receive this frame set their timers with the new node to provide decentralized synchronization. Each node uses the Q-learning algorithm to find the optimal transmission time slot. Each node acts as an agent in multi-agent reinforcement learning, where each agent faces three states. The first state is when a node starts to access the channel to send its data. The second state is when the node senses the channel and realizes it is busy but has the chance to try to access the channel and send its current data packet in the next transmission frame. The third state is when the channel is free, and the node transmits data. Time is divided using transmission frames with fixed lengths. Each transmission frame consists of 16 time slots. Each node selects between these slots, which are available actions, to send its data at the beginning of the chosen slot. The required time to ensure finishing the transmission of the current data packet in the current transmission frame is considered after the final time slot. The next transmission frame starts immediately after the current transmission frame. In Fig. 2, the structure of the transmission frame and time slots is shown.

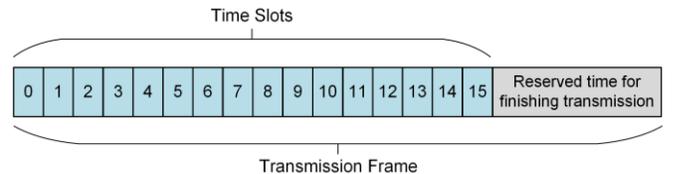

Fig. 2. The structure of the transmission frame and time slots.

At the beginning of the algorithm, each node initializes parameters. Each node uses a training data packet with the maximum possible length to find the optimal transmission slot. All nodes have the training data packet to send in all transmission frames. After the decentralized synchronization, each node uses the ϵ-greedy strategy with a decay rate to ensure a balance between exploration and exploitation and convergence to the optimal policy. For this purpose, in each iteration, a random number between 0 and 1 is generated. If ϵ is greater than this number, exploration is done by selecting a time slot randomly. If ϵ is less than the generated random number, the exploitation is done by selecting the time slot with the highest Q-value. Furthermore, $\epsilon_{min}$ is the minimum value of ϵ to ensure exploration throughout the algorithm. After selecting a time slot, the channel is checked. If the channel is busy, and the node has not reached the predefined maximum retransmission limit of the current data packet, it receives a reward equal to -10 and tries to access the channel and send data in the following transmission frame. In this case, the new state is the second state. If it reaches the maximum retransmission limit, it receives a reward equal to -20, and the transmission of the current data packet is failed. As a result, in the next transmission frame, the node should try to send its new data packet. As a result, in this case, the new state is the first state. The third situation is when the channel is free, and the node sends its data packet at the beginning of the selected time slot and receives a reward equal to +10. In this case, the new state is the third state. The value of rewards is chosen to maximize network performance, not just the performance of each node. Each node continues the Q-learning algorithm until the detection of convergence. After the convergence, each node selects its final optimal transmission time slot and sends its data packets at the beginning of the selected time slot. In the proposed algorithm, a criteria is used to detect the convergence. During the algorithm, each node considers *M* previous iterations and compares its current selected action with previously selected actions. If the current action is repeated *M-2* times in *M* previous iterations, the node detects the convergence and selects the current action as the final optimal transmission time slot. When all nodes reach the convergence, the network is converged, and all nodes send their data packets in a collision-free shared channel. In this way, the proposed algorithm provides the decentralized TDMA. The proposed algorithm is shown in Algorithm 1.

---

**Algorithm 1** Proposed algorithm

1: Initialize $Q(S_t, A_t)$, γ, α, $\epsilon_{max}$, $\epsilon_{min}$, decay rate, and *M*.
2: **if** the node is joining the network **then**
3:    Broadcast synchronization frame and set timers
4: **end if**
5: **if** a synchronization frame is received **then**
6:    Set timers
7: **end if**
8: **for** all iterations **do**
9:    Choose $A_t$ using ϵ−greedy strategy
10:   **if** convergence occurred **then**
11:       break the loop and return the final optimal transmission time slot
12:   **end if**
13:   Check the channel
14:   **if** the channel is busy **then**
15:       **if** the maximum retransmission limit exceeds **then**
16:           Receive a reward of -20 and load new data
17:       **else**
18:           Receive a reward of -10 and try to send current data in the next transmission frame
19:       **end if**
20:   **else**
21:       Transmit data at the beginning of the selected time slot and receive a reward of +10
22:   **end if**
23:   Observe $R_t$, $S_{t+1}$
24:   $Q(S_t, A_t) \leftarrow Q(S_t, A_t) + \alpha \left[ R_{t+1} + \gamma \max_a Q(S_{t+1}, a) - Q(S_t, A_t) \right]$
25:   $S_t \leftarrow S_{t+1}$
26:   **if** ϵ > $\epsilon_{min}$ **then**
27:       ϵ = ϵ ∗ (1 - decay rate)
28:   **end if**
29: **end for**

---

## 3 EXPERIMENTAL RESULTS

In this section, first, the structure of the VLC system, which is used in this paper for experiments, is discussed. Then, the experimental results show the performance of the proposed algorithm in the VLC IoT network. Average reward, convergence time, goodput, average delay, and data packet size are evaluated parameters. The proposed algorithm is compared with the CSMA/CA algorithm as a potential selection for decentralized VLC IoT networks.

### 3.1 VLC IoT network

For implementing the proposed algorithm, a VLC hardware system [9] called VLCIoT, which had been designed and implemented in our laboratory, is used. This system was designed for low data rate indoor IoT applications and operates well up to 7 m distances. In Fig. 3, the block diagram of the VLCIoT system is shown.

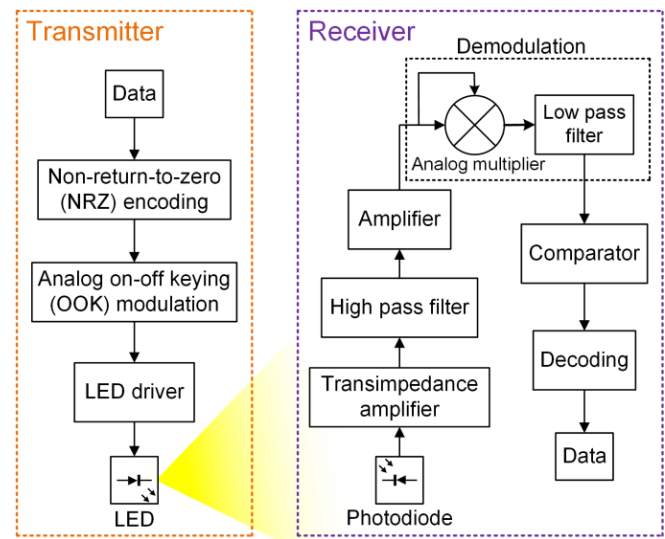

Fig. 3. Block diagram of VLCIoT system.

A VLC IoT network can be implemented using clustering. Each node collects environment parameters and sends them to the cluster head. The connection between nodes and their cluster head is based on VLC, and the connection between different cluster heads and the base station can be based on Ethernet. In this paper, only one cluster consisting of four nodes and a sink node is considered. The proposed algorithm is implemented on the microcontroller of each device using the C programming language. The network configuration is line of sight (LOS), and each node sends data to the sink node in a star topology. The semi-angle of the LED of nodes is 60°, the half-angle field of view (FOV) of the photodetector of nodes is 65°, and the angle between the transmitter and receiver of two different nodes is considered 90°. As a result, in the considered scenario, all nodes receive data from each other. Furthermore, the sink node receives data from all nodes and does not send data to them. The sink node is connected to a computer via the wire to send the received data from nodes to the computer to monitor network performance. By setting the "macRxOnWhenIdle" attribute to TRUE, the receiver of all nodes is active during idle periods. In Fig. 4, the experimental setup, and in Table 1, the selected parameters for the network and proposed algorithm are shown.

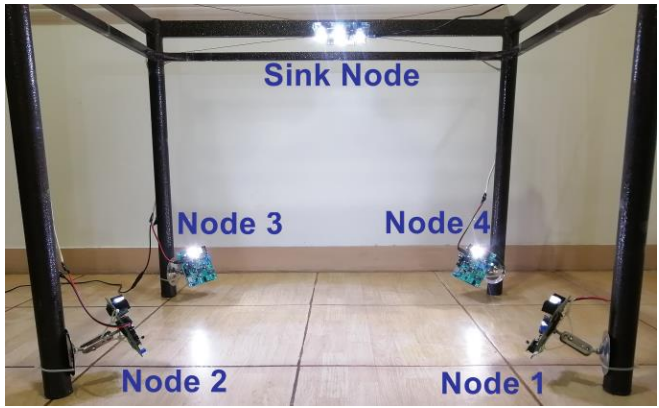

Fig. 4. The experimental setup.

Table 1. Selected Parameters for the Network and Proposed Algorithm

| Parameter | Value |
| --- | --- |
| Data rate | 115.2 kb/s |
| Training data packet size | 180 bytes |
| Transmission frame period | 100 ms |
| Time slot period | 4.6 ms |
| Number of nodes | 4 |
| Semi-angle of LEDs | 60° |
| Half-angle FOV of photodetectors | 65° |
| Maximum retransmission limit | 5 |
| *macRxOnWhenIdle* | TRUE |
| Discount factor ($\gamma$) | 0.99 |
| Learning rate ($\alpha$) | 0.2 |
| $\epsilon_{max}$ | 1 |
| $\epsilon_{min}$ | 0.05 |
| Decay rate | 0.01 |
| $M$ | 7 |

## 3.2 Learning curve

In Fig. 5, the average reward versus iterations is shown. This chart, which is called the learning curve, represents the learning process of the Q-learning algorithm in the context of transmission time slot assignments in the TDMA-based VLC IoT network. Each node in the network follows an individual learning trajectory and earns a unique average reward. In Fig. 5, the network average reward is presented, which is calculated by averaging the average reward of all nodes. By calculating the network average reward, an overview of the algorithm performance across the entire network is provided. The average reward starts from a negative value, gradually increases, and eventually converges to a positive value. This behavior can be attributed to the exploration-exploitation trade-off inherent in Q-learning. At the initial stages of learning, the algorithm explores various transmission time slot assignments, which often result in suboptimal or collision-prone actions, causing negative rewards. This exploration phase helps the algorithm gain the required knowledge about the network dynamics and discover optimal strategies. Over time, the algorithm exploits the obtained knowledge. It refines its policy, selecting transmission time slots that minimize the number of times the node faces a busy channel and maximize successful data transmission. Consequently, the average reward increases, indicating improved network performance. Eventually, the average reward converges to a positive value, indicating the convergence of the algorithm. When the convergence occurs, all nodes have found their final optimal transmission time slot. In this way, the proposed algorithm provides the decentralized TDMA.

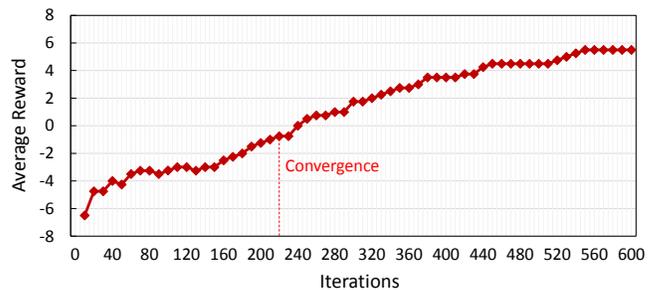

Fig. 5. The average reward versus iterations (learning curve).

In Fig. 5, the point at which the convergence occurs is shown. In the proposed algorithm, a criterion is used to detect the convergence. During the algorithm, each node considers $M$ previous iterations and compares its current selected action with previously selected actions. If the current action is repeated $M-2$ times in $M$ previous iterations, the node detects the convergence and selects the current action as the final optimal transmission time slot. When all nodes reach the convergence, the network is converged. As a result, the convergence time of the network equals the convergence time of the slowest node, shown in Fig. 5. This figure shows that the network converges at the 220th iteration. Each iteration represents a transmission frame and lasts 100 milliseconds. Therefore, the convergence time of the network is 22 seconds, which is short enough for practical VLC IoT applications. After this short time, the network is ready for collision-free data transmission using TDMA.

## 3.3 Goodput

Goodput represents the portion of time during which the network is engaged in successful data transmission [13]. In Fig. 6, the goodput of the proposed algorithm is compared with the CSMA/CA algorithm over time. For this experiment, the data packet size is 180 bytes. This chart shows the dynamic behavior of the two algorithms throughout the network operation in terms of goodput. As shown in this chart, the proposed algorithm outperforms the CSMA/CA algorithm after the convergence.

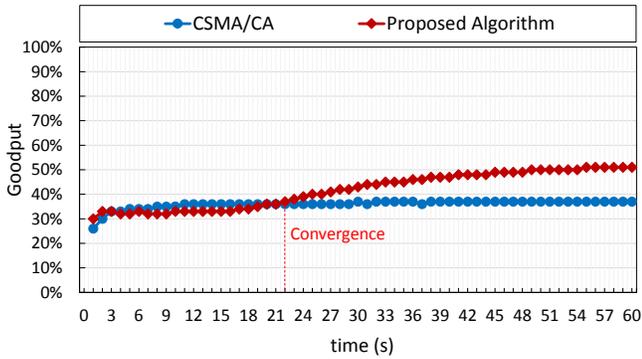

Fig. 6. The goodput versus time.

In practical VLC IoT applications, each node may be required to send different-sized data packets. To show the effect of the data packet size on the goodput, the proposed algorithm, after the convergence, is compared with the CSMA/CA. The results are shown in Fig. 7, depicting that the proposed algorithm outperforms the CSMA/CA at each of the considered packet sizes and provides up to 61% more goodput. The only drawback of the proposed algorithm regarding data packet size is that after the convergence, each node can send data packets with the maximum size of its training data packet. This limitation exists in fixed TDMA and can be solved using dynamic TDMA, which is out of the scope of this paper.

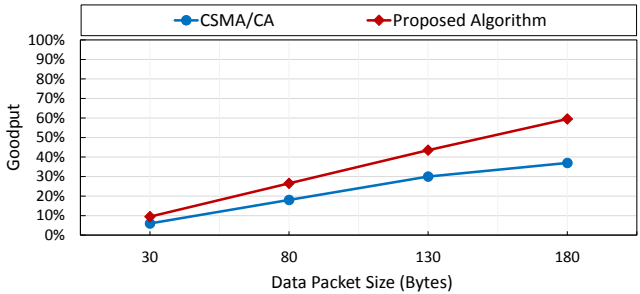

Fig. 7. The goodput versus data packet size.

## 3.4 Average delay

The average delay is defined as the average time interval between generating a data packet by a node and receiving it by the receiver node, which is the sink node in this paper, excluding the delay of the lost packets [28]. Each node in the network follows an individual learning trajectory and earns a unique average delay. In Fig. 8, the network average delay is presented, calculated by averaging the average delay of all nodes. This chart compares the average delay of the proposed algorithm with the CSMA/CA over time. For this experiment, the data packet size is 180 bytes. This chart shows the dynamic behavior of the two algorithms throughout the network operation regarding the average delay. As shown in this chart, the proposed algorithm outperforms the CSMA/CA algorithm after the convergence. In the CSMA/CA algorithm, each node may face a busy channel and postpone its transmission to the next transmission frame. The required retransmissions waste time and increase the average delay. On the other hand, after the convergence of the proposed algorithm, the collision-free TDMA ensures that each node sends its data without facing a busy channel. As a result, retransmissions are not required, and the average delay is lower than the CSMA/CA.

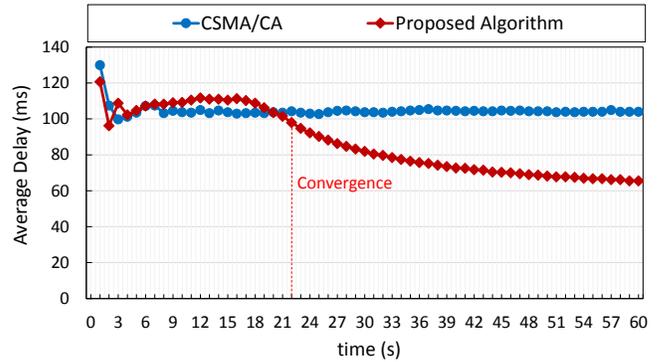

Fig. 8. The average delay versus time.

To consider the practical requirements of VLC IoT applications in terms of varying the data packet size, the proposed algorithm, after the convergence, is compared with the CSMA/CA in terms of the average delay. This comparison is shown in Fig. 9, depicting that the proposed algorithm outperforms the CSMA/CA at each of the considered packet sizes and provides up to 49% less average delay.

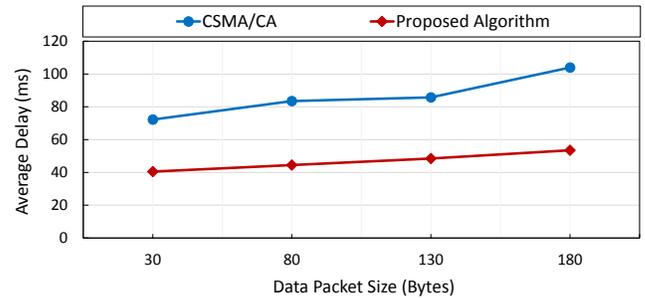

Fig. 9. The average delay versus data packet size.

## 4 CONCLUSION

This paper proposed a machine learning-based decentralized TDMA algorithm for VLC IoT networks. This paper considered a decentralized condition without a coordinator for sending synchronization frames and assigning transmission time slots to other nodes. Therefore, the traditional TDMA cannot be used. The proposed algorithm used a decentralized manner for synchronization. Furthermore, each node used the Q-learning algorithm, a reinforcement learning technique, which considered three states to find the optimal transmission time slot. Each node selects between 16 actions, which are possible transmission time slots. The proposed algorithm was implemented on a VLC hardware system, which had been designed and implemented in our laboratory. Average reward, convergence time, goodput, average delay, and data packet size were the evaluated parameters. The results showed that the proposed algorithm converges quickly and provides collision-free decentralized TDMA for the network. The proposed algorithm was compared with the CSMA/CA algorithm as a potential selection for decentralized VLC IoT networks. The results showed that the proposed algorithm provides up to 61% more goodput and up to 49% less average delay than CSMA/CA.

In this paper, the proposed algorithm assigns fixed time slots to nodes, while in some VLC IoT applications, a dynamic approach is required to change time slots for nodes over time, which will be considered for future works.

## Declaration of competing interest

The authors declare that they have no known competing financial interests or personal relationships that could have appeared to influence the work reported in this paper.

## Funding

This work was supported in part by Shahid Chamran University of Ahvaz under Grant 99/3/02/18287.